\documentstyle[11pt,newpasp,twoside]{article}

\input{epsf}

\begin{document}

\title{Gravitational Evolution of the Large-Scale Density
Distribution: The Edgeworth \& Gamma Expansions}
\author{P. Fosalba $^{1,2}$, E. Gazta\~{n}aga $^{1}$, E. Elizalde $^{1}$}
\affil{${}^1$ Institut d'Estudis Espacials de Catalunya/CSIC, Ed. Nexus-201, Gran Capit\`a 2-4, E-08034 Barcelona, Spain}
\affil{${}^2$ Astrophysics Division, Space Science Dept. of ESA/ESTEC, NL-2200 AG Noordwijk, The Netherlands}

\def\Mpc{{\,h^{-1}\,{\rm Mpc}}} 
\def\simlt{\mathrel{\spose{\lower 3pt\hbox{$\mathchar"218$}} 
     \raise 2.0pt\hbox{$\mathchar"13C$}}}

\begin{abstract}
The gravitational evolution of the cosmic 
one-point Probability Distribution Function 
(PDF)  can be estimated using an analytic approximation  
that combines gravitational Perturbation Theory (PT) 
with the Edgeworth expansion around a Gaussian PDF.
We present an alternative to the Edgeworth  
series based on an expansion around the Gamma PDF, 
which is more appropriate to describe
a realistic PDF. The Gamma expansion converges when the 
PDF exhibits exponential tails, which
are predicted by PT and N-body simulations  
in the weakly non-linear regime ({\it i.e}, 
when the variance, $\sigma^2$, is small).
We compare both expansions to 
N-body simulations and find that the Gamma expansion yields
a better overall match to the numerical results. 

\end{abstract}


\section{Introduction}

Combining non-linear perturbation theory with the
Edgeworth expansion has largely succeeded in describing
the gravitational evolution of the large-scale density PDF 
in the weakly non-linear regime, for Gaussian initial conditions
(Juszkiewicz et al 1995, Bernardeau \& Koffman 1995).
In principle, the accuracy of this approach is only limited by the
order of the (reduced) cumulants, $S_J$, involved in the Edgeworth expansion.
However, the Edgeworth series yields 
a PDF that is ill-defined. It has negative probability 
values and assigns non-zero 
probability to negative densities ($\delta<-1$).
Alternatively, we shall introduce the Gamma PDF as the basis for an 
expansion in orthogonal (Laguerre) polynomials 
around an arbitrary exponential tail 
(see Gazta\~naga, Fosalba \& Elizalde 1999).
The proposed Gamma expansion is better suited for describing a realistic PDF,
as always yields positive  
densities and the PDF is effectively positive-definite.

\section{Comparison of the expansions with N-body simulations}


\begin{figure*} 
\centering 
\centerline{\epsfxsize=7.truecm\epsfbox{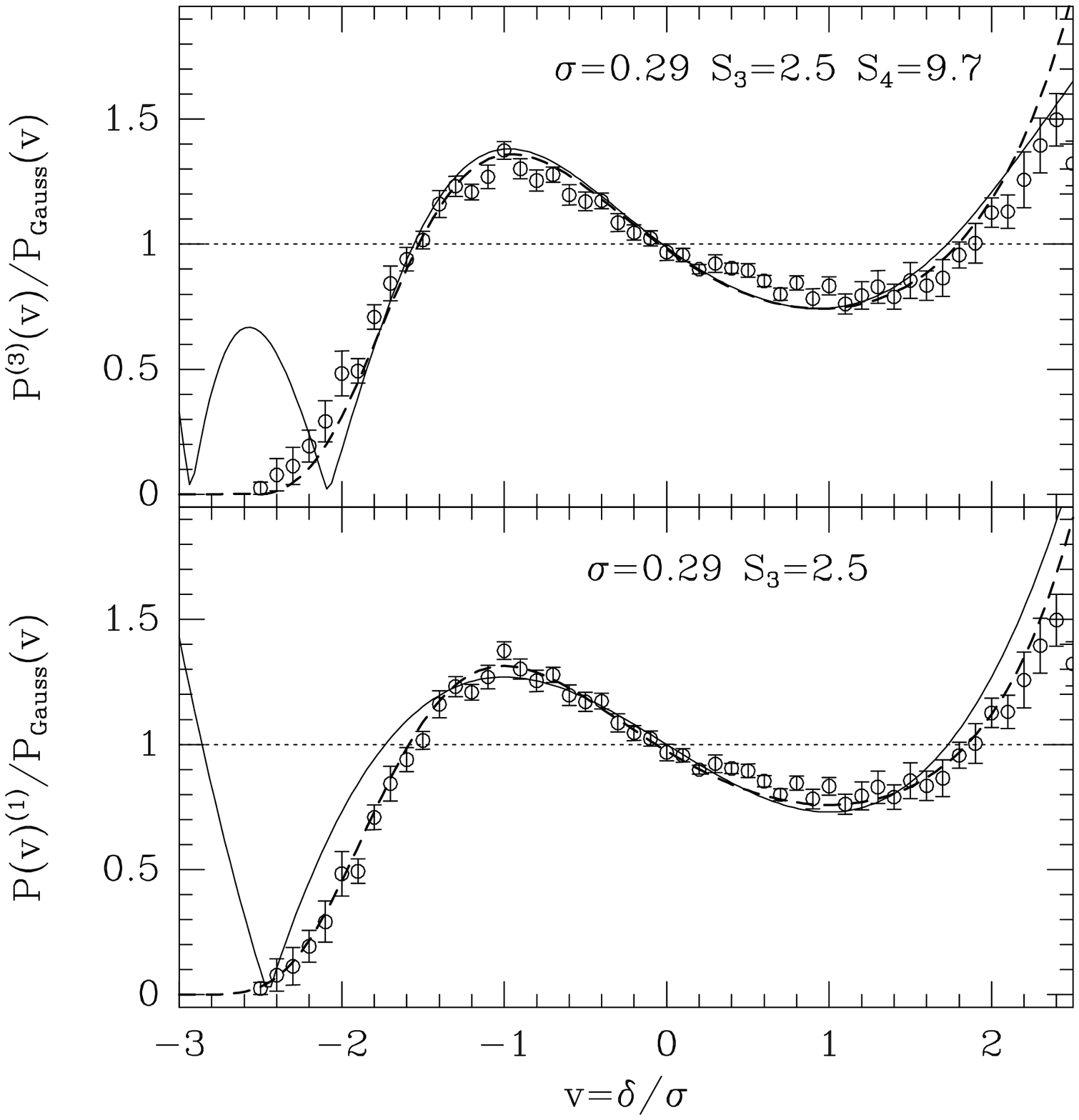}
\epsfxsize=7.truecm\epsfbox{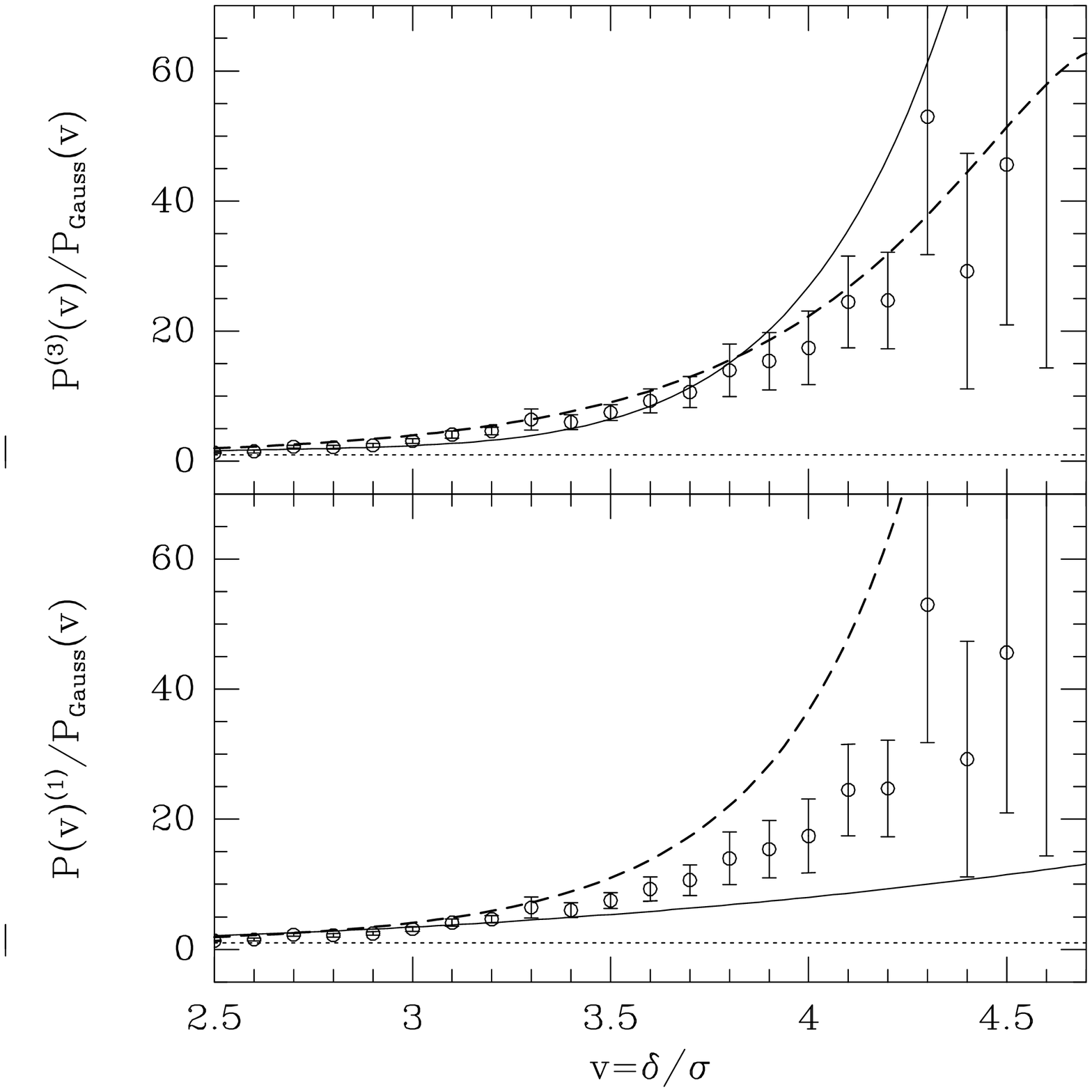}}
\caption[junk]{Deviations from the Gaussian PDF for both expansions and in
N-body simulations (symbols). 
The lower (upper) panel displays results for the first (third) order 
in powers of $\sigma$, for different values of 
the skewness $S_3$ and kurtosis $S_4$. 
The solid line is given by the Edgeworth series while the dashed one 
shows the Gamma expansion. The left and right panels show different
ranges in $\nu$.}
\label{PDFra08} 
\end{figure*}

Figure  \ref{PDFra08} shows a comparison of the Edgeworth and Gamma expansions with 
N-body simulations.
We measure the PDF in 10 realizations of  
SCDM, $\Omega=1$ and $\Gamma=0.5$,
with $L=180 \Mpc$ and $N=64^3$  
particles and $\sigma_8=1$ (Croft \& Efstathiou 1994).
We find that, up to second order, {\it both expansions 
produce very similar results, specially around the peak of the distribution},
within the error bars. 
However, 
{\it the Gamma expansion provides a better match to the PDF on the tails}.
In particular, the Gamma expansion is in far better agreement with the 
numerical results for negative values of $\nu$ (see left panel) and 
performs slightly better for the positive tail of the PDF, $\nu \simeq 1-5$
(see right panel).

In summary, we propose the Gamma expansion as a useful alternative 
to the Edgeworth series to model the gravitational evolution of the
large-scale density PDF in the weakly non-linear regime. 
We stress the potential application of the Gamma expansion 
for modeling other
non-Gaussian PDFs, such as those describing the peculiar velocities of galaxies or
the temperature anisotropy of the CMB on small scales.

\acknowledgments
This work has been 
supported by CSIC, DGICYT (Spain), project 
PB96-0925, and CIRIT (Generalitat de Catalunya), grant 1995SGR-0602.
PF is also supported by a research fellowship from ESA.

\end{document}